%% file: main.tex
\documentclass[lettersize,journal]{IEEEtran}
\usepackage{amsmath,amsfonts}
\usepackage{array}
\usepackage[caption=false,font=normalsize,labelfont=sf,textfont=sf]{subfig}
\usepackage{textcomp}
\usepackage{stfloats}
\usepackage{url}
\usepackage{verbatim}
\usepackage{graphicx}
\def\BibTeX{{\rm B\kern-.05em{\sc i\kern-.025em b}\kern-.08em
    T\kern-.1667em\lower.7ex\hbox{E}\kern-.125emX}}
\usepackage{balance}
\usepackage{glossaries}  
\usepackage{xcolor}
\usepackage{amsmath}
\usepackage{bm}
\usepackage{upgreek}
\usepackage{amssymb}
\usepackage{mathtools}
\usepackage{upgreek}
\usepackage{siunitx}
\usepackage[colorlinks=true,bookmarksopen,bookmarksnumbered,linkcolor=black,citecolor=black,urlcolor=black]{hyperref}
\loadglsentries{abbreviations.tex}

\usepackage[url=true,backend=bibtex,maxnames=99,giveninits=true]{biblatex}
\newcommand{\dt}{\ensuremath{^{n+1}}}

\begin{document}
\title{Data-driven Modified Nodal Analysis Circuit Solver}
\author{
\IEEEauthorblockN{Armin Galetzka\IEEEauthorrefmark{1}, Dimitrios Loukrezis\IEEEauthorrefmark{1}\IEEEauthorrefmark{2}, Herbert De Gersem\IEEEauthorrefmark{1}\IEEEauthorrefmark{2}} \\
\IEEEauthorblockA{\IEEEauthorrefmark{1}Institute for Accelerator Science and Electromagnetic Fields, Technische Universit{\"a}t Darmstadt, Germany}
\IEEEauthorblockA{\IEEEauthorrefmark{2}Computational Engineering, Technische Universit{\"a}t Darmstadt Darmstadt, Germany}
\thanks{Manuscript received December x, xxxx; revised August xx, xxxx. Corresponding author: A. Galetzka (email: galetzka@temf.tu-darmstadt.de).}
}

\markboth{Journal of \LaTeX\ Class Files,~Vol.~18, No.~9, September~2020}%
{How to Use the IEEEtran \LaTeX \ Templates}

\maketitle

\begin{abstract}
This work introduces a novel data-driven modified nodal analysis (MNA) circuit solver. The solver is capable of handling circuit problems featuring elements for which solely measurement data are available. Rather than utilizing hard-coded phenomenological model representations, the data-driven MNA solver reformulates the circuit problem such that the solution is found by minimizing the distance between circuit states that fulfill Kirchhoff's laws, to states belonging to the measurement data. In this way, the previously inevitable demand for model representations is abolished, thus avoiding the introduction of related modeling errors and uncertainties. The proposed solver is applied to linear and nonlinear RC-circuits and to a half-wave rectifier. 
\end{abstract}


\section{Introduction}
\input{introduction.tex}

\section{Methodology}
\label{sec:methodology}
\input{methodology.tex}

\section{Numerical Experiments}
\label{sec:numerical_experiments}
\input{numerical_experiments.tex}

\section{Computational cost}
\label{sec:computational_costs}
\input{computational_costs.tex}

\section{Conclusion}
\label{sec:conclusion}
\input{conclusion.tex}

\section*{Acknowledgment}
This work has been supported by the DFG, Research Training Group 2128 ''Accelerator Science and Technology for Energy Recovery Linacs''. 

\printbibliography

\end{document}

%% file: abbreviations.tex








\newacronym{cad}{CAD}{computer-aided design}
\newacronym[longplural=degrees of freedom]{dof}{DoF}{degree of freedom}
\newacronym{fe}{FE}{finite element}
\newacronym{fem}{FEM}{finite element method}
\newacronym{fit}{FIT}{finite integration technique}
\newacronym{nurbs}{NURBS}{non-uniform rational B-splines}
\newacronym{pec}{PEC}{perfect electric conductor}
\newacronym{pmc}{PMC}{perfect magnetic conductor}
\newacronym{rf}{RF}{radio-frequency}
\newacronym{te}{TE}{transverse electric}
\newacronym{tm}{TM}{transverse magnetic}
\newacronym{tem}{TEM}{transverse electric and magnetic}
\newacronym[longplural=boundary conditions]{bc}{BC}{boundary condition}

\newacronym{uq}{UQ}{uncertainty quantification}
\newacronym{mna}{MNA}{modified nodal analysis}
\newacronym{ode}{ODE}{ordinary differential equation}
\newacronym{map}{MAP}{maximum a posteriori probability}
\newacronym{gpr}{GPR}{Gaussian process regression}
\newacronym{pml}{PML}{perfectly matched layers}
\newacronym{fdtd}{FDTD}{finite-difference time-domain}
\newacronym{dgtd}{DGTD}{discontinuous Galerkin time-domain}
\newacronym{mim}{MIM}{metal-insulator-metal}
\newacronym{rhs}{RHS}{right-hand side}
\newacronym{sqp}{SQP}{sequential quadratic programming}
\newacronym{lar}{LAR}{least angle regression}
\newacronym{td}{TD}{total degree}
\newacronym{tp}{TP}{tensor product}
\newacronym{hc}{HC}{hyperbolic cross}
\newacronym{lhs}{LHS}{latin hypercube sampling}
\newacronym{ls}{LS}{least squares}
\newacronym{qmc}{QMC}{quasi Monte Carlo}
\newacronym{mlmc}{MLMC}{multilevel Monte Carlo}
\newacronym{hf}{HF}{high-frequency}
\newacronym{megpc}{MEGPC}{multi-element generalized polynomial chaos}
\newacronym{gp}{GP}{Gaussian process}
\newacronym{ed}{ED}{experimental design}
\newacronym{rms}{RMS}{root mean square}
\newacronym{cv}{CV}{cross-validation}
\newacronym{es}{ES}{educated sampling}
\newacronym{iga}{IGA}{isogeometric analysis}
\newacronym{bem}{BEM}{boundary element method}
\newacronym{ibvp}{IBVP}{initial/boundary value problem}


\newacronym{em}{EM}{electromagnetic}
\newacronym{emf}{EMF}{electromagnetic field}
\newacronym{gpc}{gPC}{generalized polynomial chaos}
\newacronym{mc}{MC}{Monte Carlo}
\newacronym{pde}{PDE}{partial differential equation}
\newacronym{pdf}{PDF}{probability density function}
\newacronym[longplural={quantities of interest}]{qoi}{QoI}{quantity of interest}
\newacronym{rv}{RV}{random variable}
\newacronym{ae}{a.e.}{almost every}




\DeclareMathOperator*{\argmin}{\arg\!\min}


\definecolor{TUDa-0d}{cmyk/RGB/HTML}{0,0,0,.8/83,83,83/535353}
\definecolor{TUDa-0c}{cmyk/RGB/HTML}{0,0,0,.6/137,137,137/898989}
\definecolor{TUDa-0b}{cmyk/RGB/HTML}{0,0,0,.4/181,181,181/B5B5B5}
\definecolor{TUDa-0a}{cmyk/RGB/HTML}{0,0,0,.2/220,220,220/DCDCDC}

\definecolor{TUDa-1a}{cmyk/RGB/HTML}{.7,.4,0,0/93,133,195/5D85C3}
\definecolor{TUDa-2a}{cmyk/RGB/HTML}{0.8,.2,0,0/0,156,218/009CDA}
\definecolor{TUDa-3a}{cmyk/RGB/HTML}{0.7,0,.5,0/80,182,149/50B695}
\definecolor{TUDa-4a}{cmyk/RGB/HTML}{.4,0,.8,0/175,204,80/AFCC50}
\definecolor{TUDa-5a}{cmyk/RGB/HTML}{.2,0,.8,0/221,223,72/DDDF48}
\definecolor{TUDa-6a}{cmyk/RGB/HTML}{0,.1,.7,0/255,224,92/FFE05C}
\definecolor{TUDa-7a}{cmyk/RGB/HTML}{0,.3,.8,0/248,186,60/F8BA3C}
\definecolor{TUDa-8a}{cmyk/RGB/HTML}{0,.6,.8,0 /238,122,52/EE7A34}
\definecolor{TUDa-9a}{cmyk/RGB/HTML}{0,.8,.7,0/233,80,62/E9503E}
\definecolor{TUDa-10a}{cmyk/RGB/HTML}{.2,.9,0,0/201,48,142/C9308E}
\definecolor{TUDa-11a}{cmyk/RGB/HTML}{.6,.8,0,0/128,69,151/804597}
\definecolor{TUDa-1b}{cmyk/RGB/HTML}{1,.6,0,0/0,90,169/005AA9}
\definecolor{TUDa-2b}{cmyk/RGB/HTML}{1,.3,0,0/0,131,204/0083CC}
\definecolor{TUDa-3b}{cmyk/RGB/HTML}{1,0,.6,0/0,157,129/009D81}
\definecolor{TUDa-4b}{cmyk/RGB/HTML}{.5,0,1,0/153,192,0/99C000}
\definecolor{TUDa-5b}{cmyk/RGB/HTML}{.3,0,1,0/201,212,0/C9D400}
\definecolor{TUDa-6b}{cmyk/RGB/HTML}{0,.2,1,0/253,202,0/FDCA00}
\definecolor{TUDa-7b}{cmyk/RGB/HTML}{0,.4,1,0/245,163,0/F5A300}
\definecolor{TUDa-8b}{cmyk/RGB/HTML}{0,.7,1,0/236,101,0/EC6500}
\definecolor{TUDa-9b}{cmyk/RGB/HTML}{0,1,.9,0/230,0,26/E6001A}
\definecolor{TUDa-10b}{cmyk/RGB/HTML}{.4,1,0,0/166,0,132/A60084}
\definecolor{TUDa-11b}{cmyk/RGB/HTML}{.7,1,0,0/114,16,133/721085}
\definecolor{TUDa-1c}{cmyk/RGB/HTML}{1,.7,.2,0/0,78,138/004E8A}
\definecolor{TUDa-2c}{cmyk/RGB/HTML}{1,.5,.2,0/0,104,157/00689D}
\definecolor{TUDa-3c}{cmyk/RGB/HTML}{1,.2,.6,0/0,136,119/008877}
\definecolor{TUDa-4c}{cmyk/RGB/HTML}{.6,.1,1,0/127,171,22/7FAB16}
\definecolor{TUDa-5c}{cmyk/RGB/HTML}{.4,.1,1,0/177,189,0/B1BD00}
\definecolor{TUDa-6c}{cmyk/RGB/HTML}{.2,.3,1,0/215,172,0/D7AC00}
\definecolor{TUDa-7c}{cmyk/RGB/HTML}{.2,.5,1,0/210,135,0/D28700}
\definecolor{TUDa-8c}{cmyk/RGB/HTML}{.2,.8,1,0/204,76,3/CC4C03}
\definecolor{TUDa-9c}{cmyk/RGB/HTML}{.3,1,.9,0/185,15,34/B90F22}
\definecolor{TUDa-10c}{cmyk/RGB/HTML}{.5,1,.3,0/149,17,105/951169}
\definecolor{TUDa-11c}{cmyk/RGB/HTML}{.8,1,.2,0/97,28,115/611C73}
\definecolor{TUDa-1d}{cmyk/RGB/HTML}{1,.9,.3,0/36,53,114/243572}
\definecolor{TUDa-2d}{cmyk/RGB/HTML}{1,.7,.4,0/0,78,115/004E73}
\definecolor{TUDa-3d}{cmyk/RGB/HTML}{1,.4,.7,0/0,113,94/00715E}
\definecolor{TUDa-4d}{cmyk/RGB/HTML}{.7,.3,1,0/106,139,55/6A8B22}
\definecolor{TUDa-5d}{cmyk/RGB/HTML}{.5,.2,1,0/153,166,4/99A604}
\definecolor{TUDa-6d}{cmyk/RGB/HTML}{.4,.4,1,0/174,142,0/AE8E00}
\definecolor{TUDa-7d}{cmyk/RGB/HTML}{.3,.6,1,0/190,111,0/BE6F00}
\definecolor{TUDa-8d}{cmyk/RGB/HTML}{.4,.8,1,0/169,73,19/A94913}
\definecolor{TUDa-9d}{cmyk/RGB/HTML}{.5,1,.9,0/156,28,38/961C26}
\definecolor{TUDa-10d}{cmyk/RGB/HTML}{.7,1,.5,0/115,32,84/732054}
\definecolor{TUDa-11d}{cmyk/RGB/HTML}{.9,1,.3,0/76,34,106/4C226A}

%% file: introduction.tex
\IEEEPARstart{C}{ircuit} simulations play a key role in the design and optimization process of electrical devices. To solve circuit problems, two types of equations are necessary. On the one hand, there are Kirchhoff's laws, which are derived from first principles and are accepted as exactly known \cite{Kurz2022}. On the other hand, there are model descriptions that represent the actual element behavior, which are essential in classical circuit solvers. These models are mostly empirically known, thus introducing errors and epistemic (model-form) uncertainties arising from the modeling process \cite{wit2012all}. Very commonly, these models are obtained via data-fitting techniques, e.g.,  based on physically motivated approaches \cite{shockley1949, kriplani2011_tunnel_diode} or sophisticated machine learning regression methods \cite{Humayun2010_nn_microwave, ciric2018neural, klein2022_nn}, to name a few options.

Nowadays, the availability of data increases steadily. For many circuit elements, the amount of behavioral data is unprecedentedly high, raising the opportunity for novel data-driven modeling and simulation methods. Under this light, data-driven computation has emerged as and grown into a research area of its own in recent years. 
A model-free data-driven approach was first proposed in \cite{kirchdoerfer2016data, kirchdoerfer2017data} for elasticity-related problems. Since then, the so-called data-driven computing framework has been extended to several problem classes as well \cite{Carrara2020_fracture, kirchdoerfer2018data, bai2022_dd_thin_composite, Gerbaud2022_datadriven_plasticity}, including the authors' contributions in the field of magnetic field simulation \cite{degersem2020magnetic, galetzka2020datadriven, galetzka2021datadriven}.

A common aspect in all aforementioned works is that the data-driven computing framework abolishes the previously inevitable need for model representations, e.g., regarding constitutive (material) laws. Instead, a double minimization problem is introduced to obtain the solution. Therein, the sought solution must simultaneously conform to the first-principle equations and be as close as possible to the provided measurement data. In this way, the simulation workflow is substantially reduced, as the problem at hand can be solved directly using the available measurement data, rather than investing effort in the development of model representations, which, even if considered to be adequate, remain empirical and inexact.
Contrarily, using the data-driven computing framework, the modeling process and the associated errors and uncertainties are bypassed altogether.

This paper extends data-driven computation to the case of circuit solvers. In particular, a novel data-driven \gls{mna} \cite{ho1975} solver is introduced, which is able to resolve circuit problems where only data are available for certain circuit elements. Therein, some circuit elements are assumed to be exactly known, while only measurement data are available for the remaining elements. 
The \gls{mna} is then reformulated along the lines of the data-driven computing framework, such that a double minimization solver yields the states that conform to Kirchhoff's laws, while also being as close as possible to the measurements. Three numerical experiments validate the data-driven \gls{mna} method, which is able to recover reference solutions obtained with standard \gls{mna} solvers.
As would be expected, the solutions of the data-driven \gls{mna} solver become increasingly more accurate as the number of available measurements increases.


Note that the suggested data-driven \gls{mna} solver is completely different than other approaches that appear in the literature and which are also labeled as ``data-driven'' \cite{Aadithya2020_datadriven_circuit, andrejevic2003electronic, chen2006application, chen2017verilog, gorissen2009sequential, hammouda2008neural, li2016physics, litovski1997simulation, zaabab1995neural}. Specifically, the data-driven \gls{mna} solver does not rely on machine learning regression for computing circuit element models, which can then be used within standard circuit simulation tools. Instead, the suggested data-driven \gls{mna} solution method incorporates the measurement data directly into the solver and bypasses the element modeling step altogether.

The remaining of the paper is structured as follows. In Section~\ref{sec:methodology}, we introduce the data-driven computing formulations in the \gls{mna} framework and derive the data-driven \gls{mna} circuit solver. Several numerical investigations are presented and discussed in Section~\ref{sec:numerical_experiments}, namely for a linear RC-circuit, a nonlinear RC-circuit with a voltage-dependent capacitor, and a nonlinear half-wave rectifier. A discussion on the computational cost of the data-driven \gls{mna} solver is provided in Section~\ref{sec:computational_costs}.
The paper is concluded with Section~\ref{sec:conclusion}, where we summarize our findings.

%% file: methodology.tex
\noindent In this work we consider circuits that contain passive elements, i.e., resistors (G), capacitors (C) and inductors (L), voltage sources (V) and current sources (I). Let $\mathbf{A}_X \in \mathbb{R}^{(n-1)\times n_\mathrm{X}}$ denote the reduced incidence matrices of an arbitrary circuit, where $n$ denotes the number of nodes and $X \in \{\mathrm{G}, \mathrm{C}, \mathrm{L}, \mathrm{V}, \mathrm{I}\}$. Then, employing the \gls{mna}, Kirchhoff's laws can be formulated as
\begin{subequations}
	\begin{align}
	\mathbf{A}_{\mathrm{G}}\mathbf{i}_{\mathrm{G}}(t) + \mathbf{A}_{\mathrm{C}}\dot{\mathbf{q}}_{\mathrm{C}}(t) + \mathbf{A}_{\mathrm{L}}\mathbf{i}_{\mathrm{L}}(t) + \mathbf{A}_{\mathrm{V}}\mathbf{i}_{\mathrm{V}}(t) &= \mathbf{A}_{\mathrm{I}}\mathbf{i}_{\mathrm{src}}(t), \\
	\mathbf{A}_{\mathrm{G}}^\top \boldsymbol{\Phi}(t) - \mathbf{v}_{\mathrm{G}}(t) &= 0, \\
	\mathbf{A}_{\mathrm{C}}^\top \boldsymbol{\Phi}(t) - \mathbf{v}_{\mathrm{C}}(t) &= 0, \\
	\mathbf{A}_{\mathrm{L}}^\top \boldsymbol{\Phi}(t) - \dot{\boldsymbol{\Psi}}_{\mathrm{L}}(t) &= 0, \\
	\mathbf{A}_{\mathrm{V}}^\top \boldsymbol{\Phi}(t) - \mathbf{v}_{\mathrm{src}}(t) &= 0.
	\end{align}%
	\label{eq:KCL_constraints}%
\end{subequations}%
Here, $\boldsymbol{\Phi} \in \mathbb{R}^{n-1}$ denotes the potential difference between the nodes and the mass node and $\mathbf{i}_{\mathrm{G}}\in\mathbb{R}^{n_{\mathrm{G}}},\,\mathbf{i}_{\mathrm{L}}\in\mathbb{R}^{n_{\mathrm{L}}},\,\mathbf{i}_{\mathrm{V}}\in\mathbb{R}^{n_{\mathrm{V}}},\,\mathbf{i}_{\mathrm{src}}\in\mathbb{R}^{n_{\mathrm{src}}}$ are the currents in branches with resistors, inductors, voltage sources, and current sources, respectively. The changes in charges and fluxes are denoted as $\dot{\mathbf{q}}_{\mathrm{C}}\in\mathbb{R}^{n_{\mathrm{C}}}$ and $\dot{\boldsymbol{\Psi}}_{\mathrm{L}} \in \mathbb{R}^{n_{\mathrm{L}}}$ respectively.

Applying a time-discretization scheme, e.g., the backward Euler method, the time-discrete counterpart of \eqref{eq:KCL_constraints} reads
\begin{subequations}
	\begin{align}
	\mathbf{A}_{\mathrm{G}}\mathbf{i}\dt_{\mathrm{G}} + h^{-1}\mathbf{A}_{\mathrm{C}}\mathbf{q}\dt_{\mathrm{C}}& \notag\\ + \mathbf{A}_{\mathrm{L}}\mathbf{i}\dt_{\mathrm{L}} + \mathbf{A}_{\mathrm{V}}\mathbf{i}\dt_{\mathrm{V}} &= \mathbf{A}_{\mathrm{I}}\mathbf{i}\dt_{\mathrm{src}} + h^{-1}\mathbf{A}_\mathrm{C}\mathbf{q}_\mathrm{C}^n, \\
	\mathbf{A}_{\mathrm{G}}^\top \boldsymbol{\Phi}\dt - \mathbf{v}_{\mathrm{G}}\dt &= 0, \\
	\mathbf{A}_{\mathrm{C}}^\top \boldsymbol{\Phi}\dt - \mathbf{v}_{\mathrm{C}}\dt &= 0, \\
	\mathbf{A}_{\mathrm{L}}^\top \boldsymbol{\Phi}\dt - h^{-1}\boldsymbol{\Psi}\dt_{\mathrm{L}} &= h^{-1}\boldsymbol{\Psi}_\mathrm{L}^n, \\
	\mathbf{A}_{\mathrm{V}}^\top \boldsymbol{\Phi}\dt - \mathbf{v}_{\mathrm{src}}\dt &= 0,
	\end{align}%
	\label{eq:KCL_time_discrete}%
\end{subequations}%
where $h$ denotes the time step and $f^{n+1} = f(t+h)$, respectively, $f^n = f(t)$.
To ease notation, we omit the superscript $\dt$ in the following. All states that conform to Kirchhoff's laws are collected in the set $\mathcal{K} = \{\boldsymbol{\zeta}: \boldsymbol{\zeta} = (\boldsymbol{\Phi},\mathbf{i}_\mathrm{G},\mathbf{q}_\mathrm{C},\boldsymbol{\Psi},\mathbf{i}_\mathrm{L},\mathbf{i}_\mathrm{V}) \in \mathcal{Z}: \eqref{eq:KCL_time_discrete}\}$, where $\mathcal{Z}$ denotes the set of all possible states. For a traditional circuit solver, models for the circuits elements are needed to solve \eqref{eq:KCL_time_discrete}. The models' responses can be collected in a set $\tilde{\mathcal{D}}$, which we define as
\begin{equation}
\begin{aligned}
 \tilde{\mathcal{D}} &= \left\{\bigtimes_{e} \{i = G_e(v)v\}\right\}\times \left\{\bigtimes_{e} \{ q= C_e(v)v\}\right\} \\ 
                       &\times \left\{\bigtimes_{e} \{ \Psi= L_e(i)i\}\right\},
\end{aligned}
\end{equation}
where $e$ refers to the branch that contains the considered element. The traditional solution is then determined by
\begin{equation}
	\boldsymbol{\zeta} = \mathcal{K} \cap \tilde{\mathcal{D}}.
	\label{eq:traditional_solution}
\end{equation}

The need for model representations is abolished when employing the data-driven \gls{mna} solver, within which the information on some or all circuit elements is solely available through measurement data. The per element measurement data are collected in the sets
\begin{subequations}
	\begin{align}
	\mathcal{D}_{\mathrm{G},e} &= \left\{\left(v^\dagger_{\mathrm{G}},i^\dagger_{\mathrm{G}}\right)_n\right\}_{n=1}^{N_{\mathrm{G},e}}, \\
	\mathcal{D}_{\mathrm{C},e} &= \left\{\left(v^\dagger_{\mathrm{C}},q^\dagger_{\mathrm{C}}\right)_n\right\}_{n=1}^{N_{\mathrm{C},e}}, \\
	\mathcal{D}_{\mathrm{L},e} &= \left\{\left(\Psi^\dagger_{\mathrm{L}},i^\dagger_{\mathrm{L}}\right)_n\right\}_{n=1}^{N_{\mathrm{L},e}}.
	\end{align} 
	\label{eq:measurement_data}%
\end{subequations}%
which define the global measurement set 
\begin{equation}
	\mathcal{D} = \{\bigtimes_e \mathcal{D}_{X,e}\}_{X = \{\mathrm{G},\mathrm{C},\mathrm{L}\}},
\end{equation}
where the cardinality of $\mathcal{D}$, i.e., the total number of used measurement data points, is given by $N=\sum_e N_{\mathrm{G},e} + \sum_e N_{\mathrm{C},e} + \sum_e N_{\mathrm{L},e}$.
However, searching for a solution similar to \eqref{eq:traditional_solution} yields $\boldsymbol{\zeta} = \mathcal{K} \cap {\mathcal{D}} = \emptyset$, since only a finite number of measurement data points are available. Therefore, the data-driven framework relaxes the condition \eqref{eq:traditional_solution} by accepting a solution that minimizes the distance between states that conform to Kirchhoff's laws and states belonging to the measurement set $\mathcal{D}$.
We define the distance function at each element as
\begin{subequations}
	\begin{align}
	f_{\mathrm{G},e}\left((v_{\mathrm{G}},i_{\mathrm{G}}),\mathcal{D}_{\mathrm{G},e}\right) &= \hspace{-1.4em}\min_{\substack{\phantom{A}\\(v^\dagger_{\mathrm{G}},i^\dagger_{\mathrm{G}})\in\mathcal{D}_{\mathrm{G},e}}} \hspace{-1.4em} \left\|(v_{\mathrm{G}},i_{\mathrm{G}})-(v^\dagger_{\mathrm{G}},i^\dagger_{\mathrm{G}})\right\|^2_{\tilde{G}},\\
	f_{\mathrm{C},e}\left((v_{\mathrm{C}},q_{\mathrm{C}}),\mathcal{D}_{\mathrm{C},e}\right) &= \hspace{-1.4em}\min_{\substack{\phantom{a}\\(v^\dagger_{\mathrm{C}},q^\dagger_{\mathrm{C}})\in\mathcal{D}_{\mathrm{C},e}}}\hspace{-1.4em} \left\|(v_{\mathrm{C}},q_{\mathrm{C}})-(v^\dagger_{\mathrm{C}},q^\dagger_{\mathrm{C}})\right\|^2_{\tilde{C}}, \\
	f_{\mathrm{L},e}\left((\Psi_{\mathrm{L}},i_{\mathrm{L}}),\mathcal{D}_{\mathrm{L},e}\right)  &= \hspace{-1.4em}\min_{\substack{\phantom{a}\\(\Psi^\dagger_{\mathrm{L}},i^\dagger_{\mathrm{L}})\in\mathcal{D}_{\mathrm{L},e}}} \hspace{-1.4em} \left\|(\Psi_{\mathrm{L}},i_{\mathrm{L}})-(\Psi^\dagger_{\mathrm{L}},i^\dagger_{\mathrm{L}})\right\|^2_{\tilde{L}},
	\end{align}%
	\label{eq:distance_functions}%
\end{subequations}%
with the weighted Euclidean distance $\|(\mathbf{p},\mathbf{q})\|^2_{\tilde{\mathbf{X}}} = \sum_{i=1}^M (1/2\tilde{X_i}^{-1}p_i^2 + 1/2\tilde{X_i}q_i^2)$, for  $\mathbf{p},\mathbf{q},\tilde{\mathbf{X}} \in \mathbb{R}^M$. Note that the weighting factors $\tilde{G},\,\tilde{C},\,\tilde{L}$ are only of computational nature and do not represent the underlying model of the elements \cite{kirchdoerfer2016data,galetzka2020datadriven}. For now, we only demand for constants that are bounded such that $0 < \tilde{X} < \infty$.
The global distance function is then defined as
\begin{equation}
	F(\boldsymbol{\zeta},\mathcal{D}) = F_{\mathrm{G}}\left(\mathbf{v}_{\mathrm{G}},\mathbf{i}_{\mathrm{G}}\right) + F_{\mathrm{C}}(\mathbf{v}_{\mathrm{C}},\mathbf{q}_{\mathrm{C}})+ F_{\mathrm{L}}(\mathbf{i}_{\mathrm{L}},\boldsymbol{\Psi}_{\mathrm{L}}),
\label{eq:global_distance}
\end{equation}
where the per element distance functions sum over all elements, i.e., $F_{\mathrm{X}}=\sum_{e} f_{X,e}$ for $X\in\{\mathrm{G},\mathrm{C},\mathrm{L}\}$. For a given state $\boldsymbol{\zeta}$, the distance function \eqref{eq:global_distance} returns the minimum distance of that state to the available measurement data. To obtain a solution that fulfills Kirchhoff's laws, the minimization of \eqref{eq:global_distance} simply needs to be constrained by \eqref{eq:KCL_time_discrete}, thus resulting in the constrained minimization problem
\begin{equation}
\begin{aligned}
&\min_{\boldsymbol{\zeta}} F(\boldsymbol{\zeta},\mathcal{D}) \\
&\mathrm{subject~to~\eqref{eq:KCL_time_discrete}},
\end{aligned}
\label{eq:dd_formulation}
\end{equation}
which is known as the data-driven formulation. The data-driven minimization problem \eqref{eq:dd_formulation} is essentially a double-minimization problem, which can be formulated as
\begin{equation}
\boldsymbol{\zeta} = \argmin_{\boldsymbol{\zeta}^\circ \in \mathcal{K}} \left\{\min_{\boldsymbol{\zeta}^\times \in \mathcal{D}}  \left\{ \left\| \boldsymbol{\zeta}^\circ- \boldsymbol{\zeta}^\times \right\|^2_{\tilde{G},\tilde{C},\tilde{L}}\right\}\right\},
\label{eq:double_minimization_problem}
\end{equation}
where $\|\boldsymbol{\zeta}\|^2_{\tilde{G},\tilde{C},\tilde{L}} = \|(\mathbf{v}_\mathrm{G},\mathbf{i}_\mathrm{G})\|^2_{\tilde{G}} + \|(\mathbf{v}_\mathrm{C},\mathbf{q}_\mathrm{C})\|^2_{\tilde{C}} + \|(\boldsymbol{\Psi}_\mathrm{L},\mathbf{i}_\mathrm{L})\|^2_{\tilde{L}}$.

\begin{figure}[t]
	\begin{center}
		\includegraphics[scale=0.80]{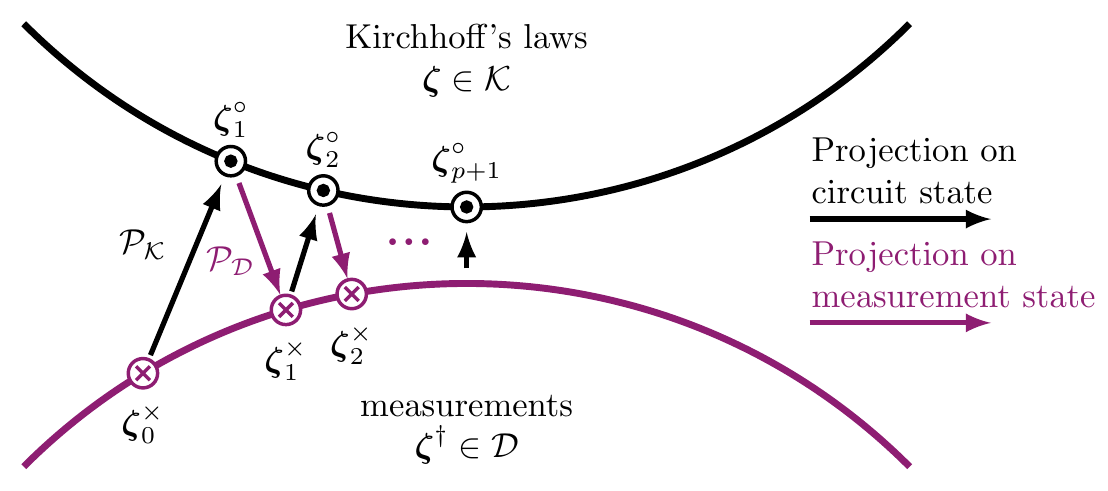}
		\caption{Illustration of the fix-point iteration.}
		\label{fig:fixpoint}
	\end{center}
\end{figure}

To solve \eqref{eq:double_minimization_problem}, we split the double minimization problem into two separate minimization problems, as follows:
\begin{enumerate}
	\item \textbf{Projection on circuit state:} Given a state $\boldsymbol{\zeta}^\times \in \mathcal{D}$, a new state $\boldsymbol{\zeta}^\circ \in \mathcal{K}$ which is compatible with Kirchhoff's laws and closest to a measurement state is found by solving
	\begin{equation}
	\begin{aligned}
	&\boldsymbol{\zeta}^\circ = \argmin_{\boldsymbol{\zeta}^\circ \in \mathcal{K}} \left\| \boldsymbol{\zeta}^\circ- \boldsymbol{\zeta}^\times \right\|^2_{\tilde{G},\tilde{C},\tilde{L}}, \\
	&\mathrm{subject~to~\eqref{eq:KCL_time_discrete}}.
	\end{aligned}
	\label{eq:min_towards_measurement}
	\end{equation}
	The constrained minimization problem can be solved with the help of Lagrange multipliers and by replacing the unknown voltages by potentials, which yields
	\begin{equation}
	\begin{aligned}
	\mathcal{L} = &\left\|\left(\mathbf{A}_{\mathrm{G}}^\top \boldsymbol{\Phi}, \mathbf{i}_{\mathrm{G}}\right)- \left(\mathbf{v}_{\mathrm{G}}^\times,\mathbf{i}_{\mathrm{G}}^\times\right)\right\|_{\tilde{G}}^2 +
	\\
	&\left\|\left(\mathbf{A}_{\mathrm{C}}^\top \boldsymbol{\Phi}, \mathbf{q}_{\mathrm{C}}\right)- \left(\mathbf{v}_{\mathrm{C}}^\times,\mathbf{q}_{\mathrm{C}}^\times\right)\right\|_{\tilde{C}}^2 + \\
	&\left\|\left(\boldsymbol{\Psi}_{\mathrm{L}}, \mathbf{i}_{\mathrm{L}}\right)- \left(\boldsymbol{\Psi}_{\mathrm{L}}^\times,\mathbf{i}_{\mathrm{L}}^\times\right)\right\|_{\tilde{L}}^2 - \\
	&\boldsymbol{\upeta}^\top \left(\mathbf{A}_{\mathrm{G}}\mathbf{i}_{\mathrm{G}}+\mathbf{A}_{\mathrm{C}}\frac{\mathbf{q}_{\mathrm{C}}- \mathbf{q}_{\mathrm{C}}^n}{h}\right) - \\
	&\boldsymbol{\upeta}^\top\left(\mathbf{A}_{\mathrm{L}}\mathbf{i}_{\mathrm{L}}+\mathbf{A}_{\mathrm{V}}\mathbf{i}_{\mathrm{V}}+\mathbf{A}_{\mathrm{I}}\mathbf{i}_{\mathrm{src}} \right)- \\
	&\boldsymbol{\uplambda}_{\mathrm{L}}^\top \left(\mathbf{A}_{\mathrm{L}}^\top \boldsymbol{\Phi} -  \frac{\boldsymbol{\Psi}_{\mathrm{L}}- \boldsymbol{\Psi}_{\mathrm{L}}^n}{h}\right) - \\
	&\boldsymbol{\uplambda}_{\mathrm{V}}^\top \left(\mathbf{A}_{\mathrm{V}}^\top \boldsymbol{\Phi} - \mathbf{v}_{\mathrm{src}}\right).
	\end{aligned}
	\label{eq:Lagrangian}
	\end{equation}
The minimum of \eqref{eq:Lagrangian} with respect to $\boldsymbol{\Phi},\mathbf{i}_\mathrm{G},\mathbf{q}_\mathrm{C},\boldsymbol{\Psi},\mathbf{i}_\mathrm{L},\mathbf{i}_\mathrm{V}$ is found by taking the derivatives with respect to the unknowns, which results in a linear system of the form $\mathbf{M}\boldsymbol{\zeta} = \mathbf{b}$. The system matrix $\mathbf{M}$ is given by
\begin{equation}
\hspace{-0.5em}
\begin{bsmallmatrix}
\mathbf{A}_{\mathrm{G}} \tilde{\mathbf{G}}\mathbf{A}_{\mathrm{G}}^\top \\ +\mathbf{A}_{\mathrm{C}}\tilde{\mathbf{C}}\mathbf{A}_{\mathrm{C}}^\top & 0            &  0        &  0           & 0            &    0         & 0                  &   -\mathbf{A}_{\mathrm{L}}   &    -\mathbf{A}_{\mathrm{V}}\\   
0 & \tilde{\mathbf{R}}    &  0        &  0           & 0            &    0         & - \mathbf{A}_{\mathrm{G}}^\top &   0               &    0\\   
0 & 0            & \tilde{\mathbf{K}} &  0           & 0            &    0         & -h^{-1}\mathbf{A}_{\mathrm{C}}^\top          &   0               &    0\\   
0 & 0            &  0        & \tilde{\mathbf{L}}    & 0            &    0         & -\mathbf{A}_{\mathrm{L}}^\top  &   0               &    0\\   
0 & 0            &  0        & 0            & \tilde{\mathbf{M}}    &    0         & 0                  &   h^{-1}                 &    0\\   
0 & 0            &  0        & 0            & 0            &    0         & -\mathbf{A}_{\mathrm{V}}^\top &   0               &    0\\   
0 & \mathbf{A}_{\mathrm{G}} &  h^{-1}\mathbf{A}_{\mathrm{C}}      &  \mathbf{A}_{\mathrm{L}} & 0            & \mathbf{A}_{\mathrm{V}} & 0                  &   0               &    0\\   
\mathbf{A}_{\mathrm{L}}^\top & 0            &  0        & 0            & -h^{-1}            &    0         & 0                  &   0               &    0\\   
\mathbf{A}_{\mathrm{V}}^\top & 0            &  0        & 0            & 0            &    0         & 0                  &   0               &    0\\   
\end{bsmallmatrix}
\label{eq:system_matrix}
\end{equation}
and the right hand side reads
\begin{equation}
	 \mathbf{b} = 
	\begin{bsmallmatrix}
	\mathbf{A}_\mathrm{G} \tilde{\mathbf{G}} \mathbf{v}_\mathrm{G}^\times + \mathbf{A}_\mathrm{C}\tilde{\mathbf{C}}\mathbf{v}_\mathrm{C}^\times \\
	\tilde{\mathbf{R}}\mathbf{i}_\mathrm{G}^\times \\
	\tilde{\mathbf{K}} \mathbf{q}_\mathrm{C}^\times \\
	\tilde{\mathbf{L}}\mathbf{i}_\mathrm{L}^\times \\
	\tilde{\mathbf{M}}\boldsymbol{\Psi}_\mathrm{L}^\times \\
	0 \\
	h^{-1} \mathbf{A}_\mathrm{C}\mathbf{q}_\mathrm{C}^n-  \mathbf{A}_\mathrm{I} \mathbf{i}_\mathrm{src} \\
	-h^{-1} \boldsymbol{\Psi}_\mathrm{L}^n \\
	\mathbf{v}_\mathrm{src}
	\end{bsmallmatrix}.
	\label{eq:rhs}
\end{equation}
Thus, a state $\boldsymbol{\zeta}^\circ \in \mathcal{K}$ that is closest to a given state $\boldsymbol{\zeta}^\times \in \mathcal{D}$ is obtained with
\begin{equation}
	\boldsymbol{\zeta}^\circ = \mathcal{P}_\mathcal{K}(\boldsymbol{\zeta}^\times),
\end{equation}
where $\mathcal{P}_\mathcal{K}$ denotes the closest point projection of a state $\boldsymbol{\zeta}^\times \in \mathcal{D}$ to $\mathcal{K}$.

\item \textbf{Projection on measurement state:} Given a state $\boldsymbol{\zeta}^\circ \in \mathcal{K}$, a new state $\boldsymbol{\zeta}^\times \in \mathcal{D}$ which belongs to the measurement set and is closest to $\boldsymbol{\zeta}^\circ$ is found by solving
\begin{equation}
\boldsymbol{\zeta}^\times = \argmin_{\boldsymbol{\zeta}^\times \in \mathcal{D}} \left\| \boldsymbol{\zeta}^\circ- \boldsymbol{\zeta}^\times \right\|^2_{\tilde{G},\tilde{C},\tilde{L}}.
\label{eq:min_towards_KL}
\end{equation}
Problem \eqref{eq:min_towards_KL} is essentially a nearest neighbor problem, which can be solved with brute-force methods or kd-tree methods, to name but a few options \cite{taunk2019brief}. Note that the minimization problem \eqref{eq:min_towards_KL} is solved independently for each element. Introducing the operator $\mathcal{P}_\mathcal{D}$ that denotes the closest point projection of a state $\boldsymbol{\zeta}^\circ \in \mathcal{K}$ to $\mathcal{D}$, a new state $\boldsymbol{\zeta}^\times \in \mathcal{D}$ is obtained through
\begin{equation}
	\boldsymbol{\zeta}^\times = \mathcal{P}_\mathcal{D}(\boldsymbol{\zeta}^\circ).
\end{equation}
\end{enumerate}
Thus for each solver iteration, the two minimization problems  \eqref{eq:min_towards_measurement} and \eqref{eq:min_towards_KL} must be solved. Depending on the initial value, the iteration scheme can be defined as
\begin{equation}
	\boldsymbol{\zeta}^\circ_{p+1} = \left(\mathcal{P}_\mathcal{M}\circ \mathcal{P}_\mathcal{D}\right)(\boldsymbol{\zeta}^\circ_p),
\end{equation}
respectively
\begin{equation}
\boldsymbol{\zeta}^\times_{p+1} = \left(\mathcal{P}_\mathcal{D}\circ \mathcal{P}_\mathcal{M}\right)(\boldsymbol{\zeta}^\times_p),
\end{equation}
which can be seen as a fix-point iteration, as visually illustrated in Figure~\ref{fig:fixpoint}. 
This minimization procedure is carried out until either $\boldsymbol{\zeta}^\circ_{p}$ and $\boldsymbol{\zeta}^\times_p$ do not change after two consecutive iterations or until a desired accuracy is reached. The accuracy at iteration $p$ can be measured in terms of the energy mismatch $\epsilon_ {\mathrm{em},p}$, which is defined as
\begin{equation}
	\epsilon_{\mathrm{em},p}^2 = \left\| \boldsymbol{\zeta}^\circ_p- \boldsymbol{\zeta}^\times_p \right\|^2_{\tilde{G},\tilde{C},\tilde{L}},
	\label{eq:energy_mismatch}
\end{equation}
which is essentially the weighted distance between the two states. Convergence is then achieved if the change in energy mismatch \eqref{eq:energy_mismatch} between two consecutive iterations is below a user-defined tolerance. Once the data-driven solver has converged, we carry on with the next time step. The most recent data-driven solution is employed as starting point in the new time step.

It is customary that most of the elements in a circuit are well-known and model representations are available. In this case, various approaches combining exactly known models with data-driven computing have been proposed by the authors in \cite{degersem2020magnetic}. We consider the least intrusive approach, where the minimization \eqref{eq:min_towards_KL} is now performed to the known model representation. In addition, the weighting factors in \eqref{eq:system_matrix} and \eqref{eq:rhs} are chosen according to the known model coefficients.
Furthermore, it has been shown by the authors that, for elements featuring a strongly nonlinear response and where only data-starved or unevenly filled measurement sets are available, constant weighting factors hinder the convergence rate of the data-driven solver \cite{galetzka2020datadriven}. Therefore, we follow the approach developed in \cite{galetzka2020datadriven} and choose the weighting factors to be the local tangent of the current state with respect to the surrounding states in the measurement set.

%% file: numerical_experiments.tex
In the following, we consider different circuit configurations. To validate the data-driven solver, we generate artificial measurement data of increasing cardinality, for each of which the circuit is solved. At every time step $t_k,\,k=1,\dots,K$, and at a specific 
element $X\in\{\mathrm{G},\mathrm{C},\mathrm{L}\}$, we calculate the energy mismatch with respect to an analytic reference solution, e.g., obtained with a traditional \gls{mna} solver utilizing the ``true'' circuit elements. 
The energy mismatch error is defined as
\begin{equation}
	\epsilon_{\mathrm{EM},\tilde{X}}^2(t) = \left.\left\|(p_\mathrm{ref},q_\mathrm{ref})-(p,q) \right\|^2_{\tilde{X}}\right|_t.
	\label{eq:rel_energy_mismatch}
\end{equation}
Furthermore, we define the \gls{rms} error over the entire time interval as
\begin{equation}
\begin{aligned}
\epsilon_{\mathrm{RMS},\tilde{X}} &= \left(\frac{\int_{t_0}^T \epsilon_{\mathrm{EM},\tilde{X}}^2(t)\,\mathrm{d}t}{\int_{t_0}^T (\|(p_\mathrm{ref},q_\mathrm{ref})\|^2_{\tilde{X}} \,\mathrm{d}t}\right)^{\frac{1}{2}} \\ &
\approx  \left(\frac{\sum_{k=0}^{K}\epsilon_{\mathrm{EM},\tilde{X}}^2(t_k)}{\sum_{k=0}^{K}(\|(p_\mathrm{ref},q_\mathrm{ref})\|^2_{\tilde{X}})|_{t_k}}\right)^{\frac{1}{2}},
\end{aligned}
	\label{eq:rms_over_time}
\end{equation} 
The weighting factors in \eqref{eq:rel_energy_mismatch} and \eqref{eq:rms_over_time} correspond to the actual model parameters which are used for the reference solution.

\subsection{Linear RC-circuit}
\label{subsec:linear-RC-circuit}
For a first test, we consider an RC-circuit, with R and C in series. The circuit is excited by a constant voltage source. 
Furthermore, in order to use the analytical solution as a reference, we consider only linear elements. 
We employ a data-driven solver based on the backward Euler scheme and on the trapezoidal rule, where the resistor and the capacitor are assumed to be unknown and only measurement data is available for these elements.
Additionally, solutions for different time step values, equivalently, numbers of time steps, are computed. 

\begin{figure}[t]
	\begin{center}
		\includegraphics[scale=0.80]{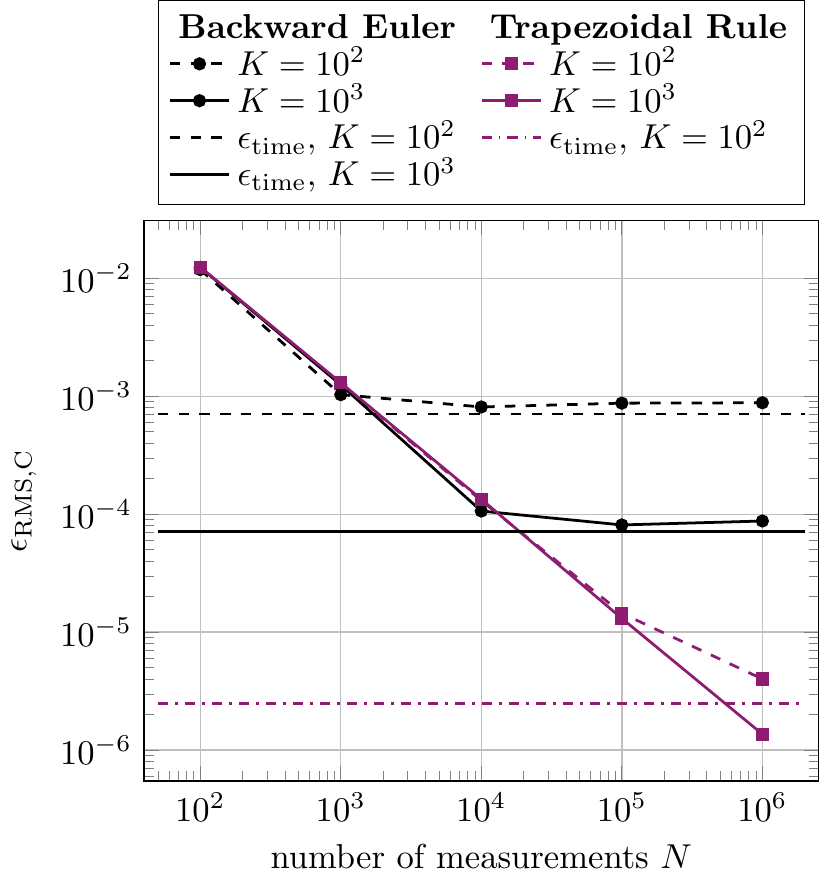}
		\caption{Linear RC-circuit: \gls{rms} error at the capacitor C over number of measurements. The horizontal lines mark the time-discretization errors for a traditional \gls{mna} circuit solver.}
		\label{fig:RC_errRMS}
	\end{center}
\end{figure}

Figure~\ref{fig:RC_errRMS} shows the \gls{rms} error for solutions obtained with the backward Euler method and the trapezoidal rule for different numbers of time steps. 
Furthermore, the \gls{rms} error for a traditional \gls{mna} solver is shown.
The results show that the accuracy of the data-driven solution improves as more measurement data are employed. However, for certain combinations of employed measurement data and number of time steps, the convergence stagnates. To analyse this behavior, we consider the overall error of a state $(p,q)$ 
to the reference solution. The error is given as 
\begin{equation}
\begin{aligned}
\epsilon &= \left\|(p_\mathrm{ref},q_\mathrm{ref})-(p,q) \right\|^2_{\tilde{X}} \\
&=\left\|(p_\mathrm{ref}-p_\mathrm{trad},q_\mathrm{ref}-q_\mathrm{trad})-(p-p_\mathrm{trad},q-q_\mathrm{trad}) \right\|^2_{\tilde{X}} \\
&\le \underbrace{\left\|(p_\mathrm{ref},q_\mathrm{ref})-(p_\mathrm{trad},q_\mathrm{trad}) \right\|^2_{\tilde{X}}}_{\epsilon_\mathrm{time}} \\
&+ \underbrace{\left\|(p,q)-(p_\mathrm{trad},q_\mathrm{trad}) \right\|^2_{\tilde{X}}}_{\epsilon_{\mathrm{data}}},
\end{aligned}
\end{equation}
where $(p_\mathrm{trad},q_\mathrm{trad})$ is a state computed with a traditional \gls{mna} solver. Hence, the overall error can be decomposed into 
an error stemming from the time discretization scheme and an error attributed to the finite measurement data. This decomposition can be clearly observed in Figure~\ref{fig:RC_errRMS}. The figure shows several cases where extending the measurement set does not lead to an improvement in the accuracy. In those 
cases, the time-discretization error $\epsilon_\mathrm{time}$, is dominant and the accuracy of the data-driven solver stagnates at the accuracy level of the traditional solver. Hence, to avoid unnecessary computational costs, the time discretization scheme and the number of 
employed measurement data points should preferably be chosen according to one another. In cases where the time discretization error $\epsilon_\mathrm{time}$ is negligible, e.g., for $K=1000$ and the trapezoidal rule, the data-driven solver achieves a linear convergence rate with respect to the employed measurement data.

\begin{figure}[t!]
	\begin{center}
		\includegraphics[scale=0.80]{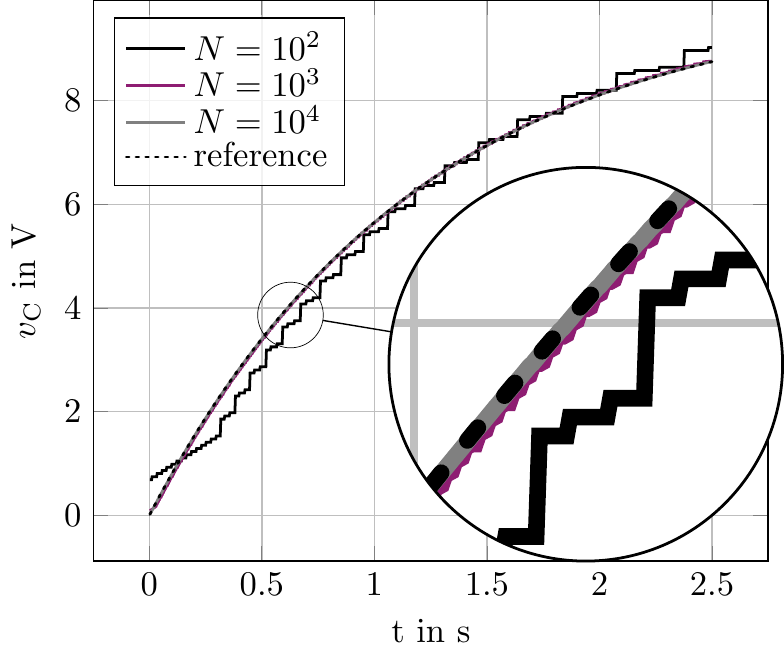}
		\caption{Linear RC-circuit: Voltage $v_\mathrm{C}$ at the capacitor over time. The plot shows the analytical reference solution, as well as data-driven solutions for an increasing number of measurement data.}
		\label{fig:RC_uC_over_time}
	\end{center}
\end{figure}

\begin{figure}[t!]
	\begin{center}
		\includegraphics[scale=0.80]{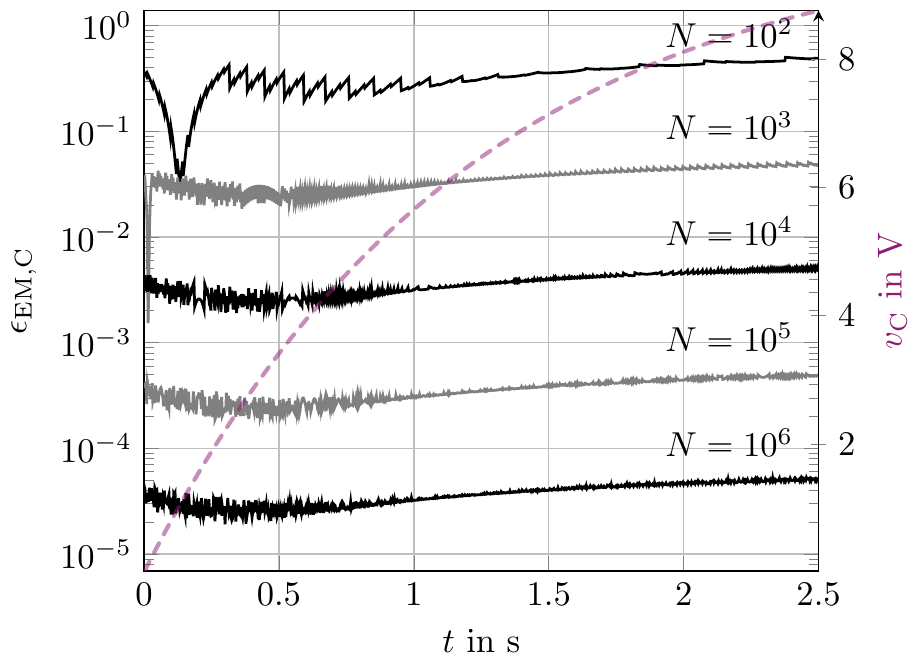}
		\caption{Linear RC-circuit: Energy mismatch at the capacitor C over time. The error is shown for an increasing number of measurement data $N \in \left\{10^2, 10^3, 10^4, 10^5, 10^6\right\}$. The dashed line shows the voltage $v_\mathrm{C}$.}
		\label{fig:RC_errRMS_over_time}
	\end{center}
\end{figure}

Next, we focus on the error attributed to the limited measurement data. Therefore, the circuit is simulated using the the trapezoidal rule and $K=1000$ time steps.
The reference solution $v_\mathrm{C}$ as well as data-driven solutions for measurement data sets of increasing size are depicted in Figure~\ref{fig:RC_uC_over_time}. We can already qualitatively observe that, if the measurement data set is not sufficiently large, e.g., for $N=10^2$, the solution of the data-driven solver captures the behavior of the circuit, albeit a large error to the analytic 
solution remains. Furthermore, the figure shows that due to the discrete data, the solution is not continuous and exhibits jumps at certain time steps. More precisely, due to the finite data, the solver might select the same measurement state in consecutive time steps. However, as more data become available, these 
discontinuities are reduced. Figure~\ref{fig:RC_errRMS_over_time} shows the energy mismatch to the reference solution over time for measurement sets of 
increasing cardinality. The figure clearly shows that the solution accuracy increases with the number of measurement data points. 

\subsection{Nonlinear RC-circuit}
\label{subsec:nonlinear-RC-circuit}
\begin{figure}[t!]
	\begin{center}
		\includegraphics[scale=0.80]{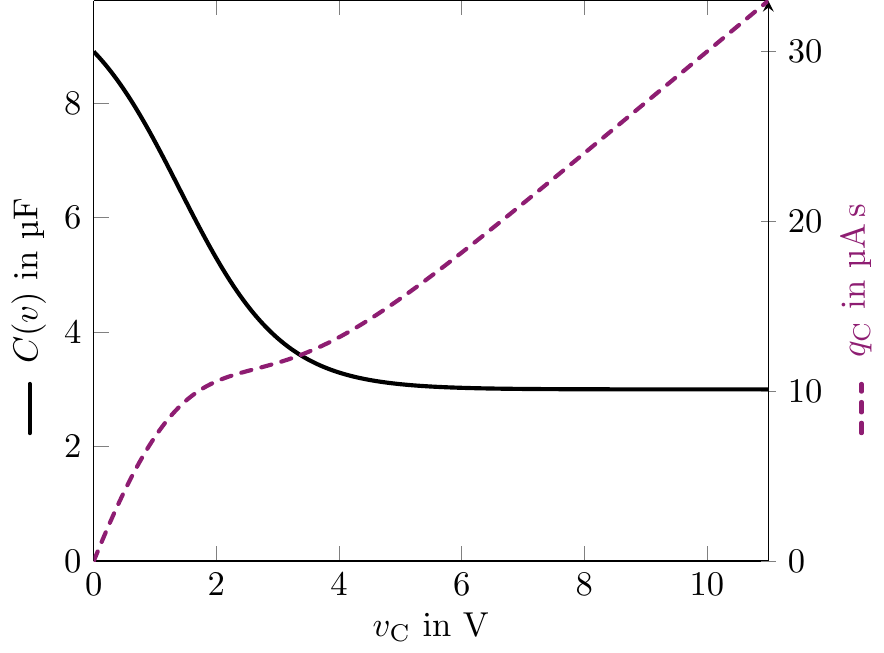}
		\caption{Nonlinear RC-circuit: Voltage-dependent capacitor $C(v_\mathrm{C})$ and charge $q_\mathrm{C}(v_\mathrm{C})$ over applied voltage $v_\mathrm{C}$.}
		\label{fig:nonlinear_capacitor}
	\end{center}
\end{figure}

We revisit the RC-circuit, yet this time we consider a voltage-dependent capacitor $C(v_\mathrm{C}(t))$. The response of the capacitor as well as the charge over the applied voltage is shown in Figure~\ref{fig:nonlinear_capacitor}. This nonlinear response is the typical behavior of multi-layer ceramic capacitors, for which the capacitance drops rapidly for moderate voltages to a nearly constant value \cite{Scheier2013_MLC,Zeltser2018_MLC}.

\begin{figure}[t!]
	\begin{center}
		\includegraphics[scale=0.80]{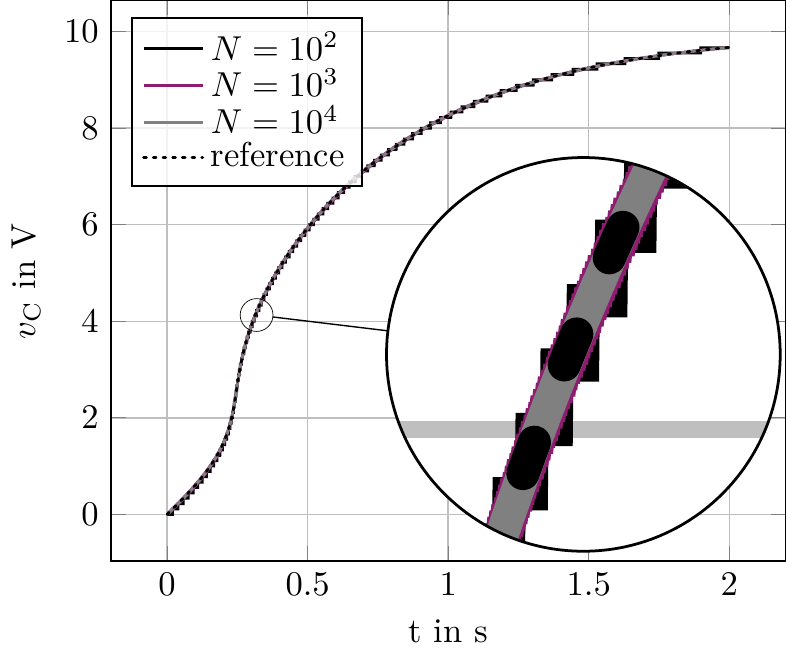}
		\caption{Nonlinear RC-circuit: voltage $v_\mathrm{C}$ at the capacitor over time. The plot shows the analytical reference solution, as well as data-driven solutions for a increasing number of measurement data $N \in \left\{10^2, 10^3, 10^4\right\}$.}
		\label{fig:RC_nonlin_uC_over_time}
	\end{center}
\end{figure}

\begin{figure}[t!]
	\begin{center}
		\includegraphics[scale=0.80]{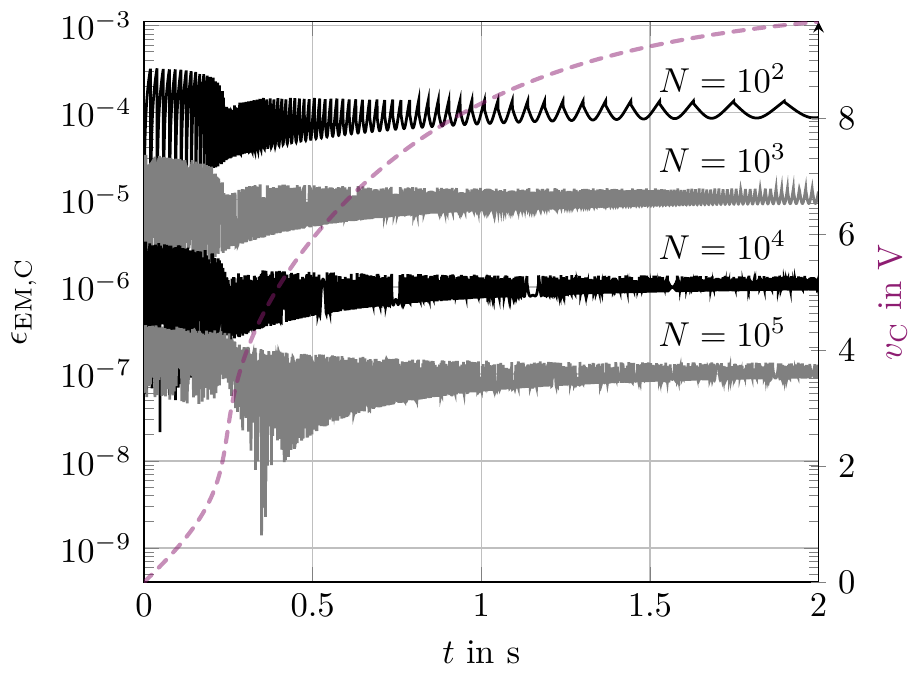}
		\caption{Nonlinear RC-circuit: Energy mismatch at the capacitor C over time. The error is shown for an increasing number of measurement data $N \in \left\{10^2, 10^3, 10^4, 10^5\right\}$. The dashed line shows the voltage $v_\mathrm{C}$.}
		\label{fig:RC_nonlin_errC_over_time}
	\end{center}
\end{figure}

\begin{figure}[t!]
	\begin{center}
		\includegraphics[scale=0.80]{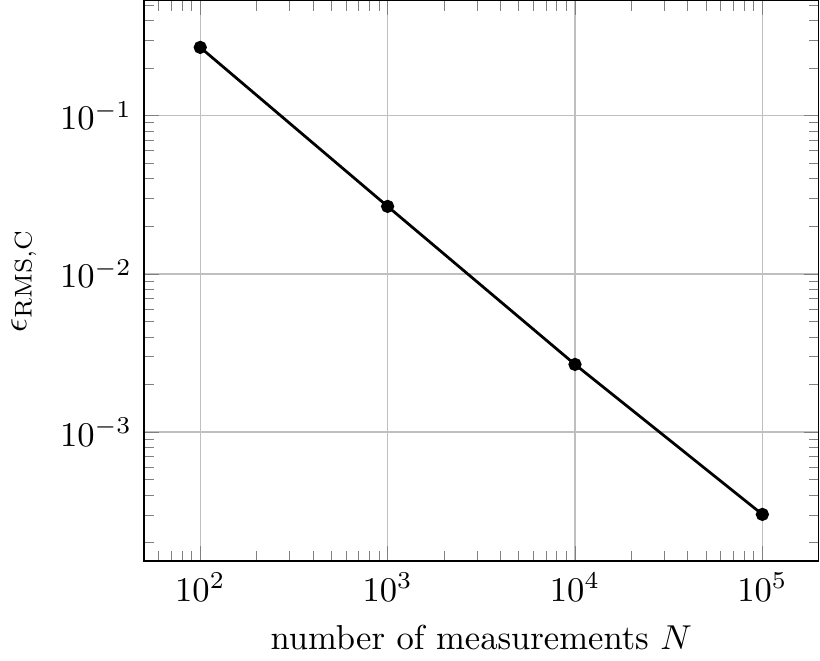}
		\caption{Nonlinear RC-circuit: \gls{rms} error at the capacitor C over number of measurements.}
		\label{fig:RC_nonlin_conv_RMS}
	\end{center}
\end{figure}

The data-driven solver treats the resistor as a known element. Contrarily, only measurement data is available for the voltage-dependent capacitor. To validate the results obtained with the data-driven solver, we generate measurement data sets of increasing cardinality. Both the reference solver and the data-driven solver employ the trapezoidal rule for time discretization. Figure~\ref{fig:RC_nonlin_uC_over_time} shows the capacitor voltage over time. Similar to the linear case, the data-driven solution shows large jumps between the time steps if the number of measurement data points is not sufficiently large. As the number of measurements increases, the data-driven solution becomes increasingly smoother and eventually identical to the conventional \gls{mna} solution. This is also evident by the results shown in Figure~\ref{fig:RC_nonlin_conv_RMS} regarding the \gls{rms} error over time for an increasing number of measurement data points. Similar to the linear RC circuit, a linear convergence rate with respect to the number of measurement points is observed.
The energy mismatch over time and for measurement sets of increasing cardinality is shown in Figure~\ref{fig:RC_nonlin_errC_over_time}. Again, the figure clearly shows that the solution accuracy increases with the number of measurement data points. We also observe that the error in time is slightly larger in the strongly nonlinear region, that is, for $t< \SI{0.5}{\second}$, respectively for $v_\mathrm{C}<\SI{4}{\volt}$.

\subsection{Half-wave rectifier}
\label{subsec:rectifier}
\begin{figure}[t!]
	\begin{center}
		\includegraphics[scale=0.80]{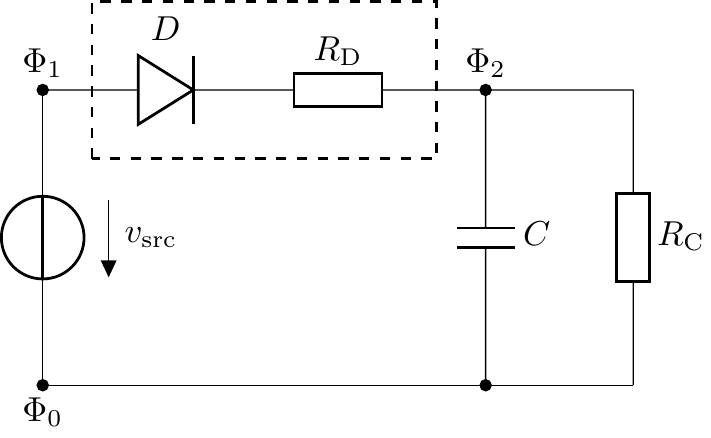}
		\caption{Half-wave rectifier circuit.}
		\label{fig:half-wave-rectifier}
	\end{center}
\end{figure}
The next test case considers the well-known half-wave rectifier circuit, depicted in Figure~\ref{fig:half-wave-rectifier}. The circuit elements are $R_\mathrm{C}=\SI{1}{\kilo\ohm}$, $R_\mathrm{D}=\SI{10}{\milli\ohm}$, $C=\SI{100}{\micro\farad}$ and $v_\mathrm{src} = \SI{5}{\volt}\sin(\omega t)$, where $\omega = 2\pi f$ and $f=\SI{100}{\hertz}$. The diode is modeled using the Shockley diode model \cite{shockley1949}, such that
\begin{equation}
i(t) = i_\mathrm{s} \left( \mathrm{e}^{\frac{v_\mathrm{D}(t)}{n v_\mathrm{T}}} - 1\right),
\label{eq:shockley_diode}
\end{equation}
where $v_\mathrm{D}$ is the voltage at the diode, $i_\mathrm{s}=\SI{2.52}{\nano\ampere}$ is the saturation current, $v_\mathrm{T}=\SI{25.85}{\milli\volt}$ is the thermal voltage, here for $T=\SI{300}{\kelvin}$, and $n=1.752$ is the ideality factor. 

\begin{figure}[t!]
	\begin{center}
		\includegraphics[scale=0.80]{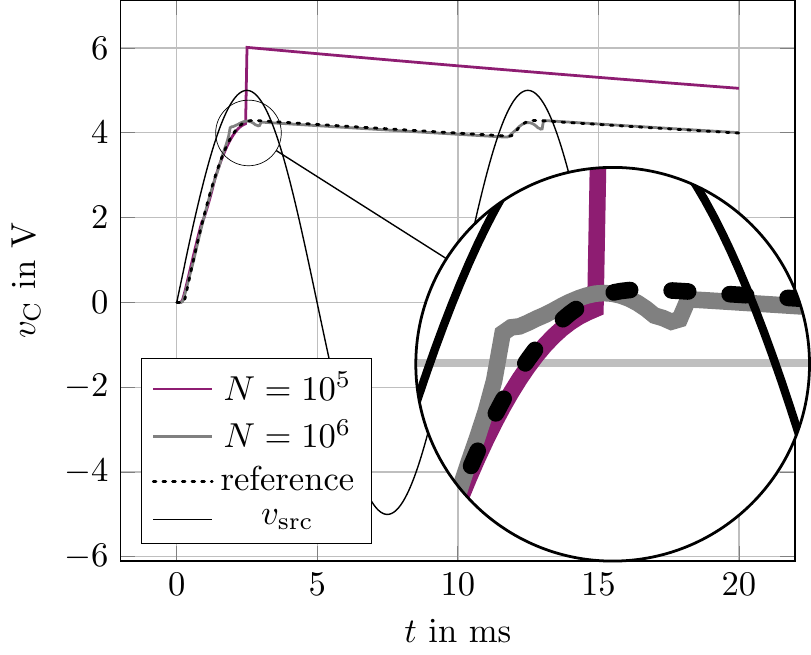}
		\caption{Half-wave rectifier: Voltage $v_\mathrm{C}$ at the capacitor over time. The plot shows the reference solution obtained with \textsc{ngspice} and data-driven solutions for an increasing number of measurement data $N \in \left\{10^5, 10^6\right\}$.}
		\label{fig:half-wave-rectifier_uC}
\end{center}
\end{figure}

\begin{figure}[t!]
	\begin{center}
		\includegraphics[scale=0.80]{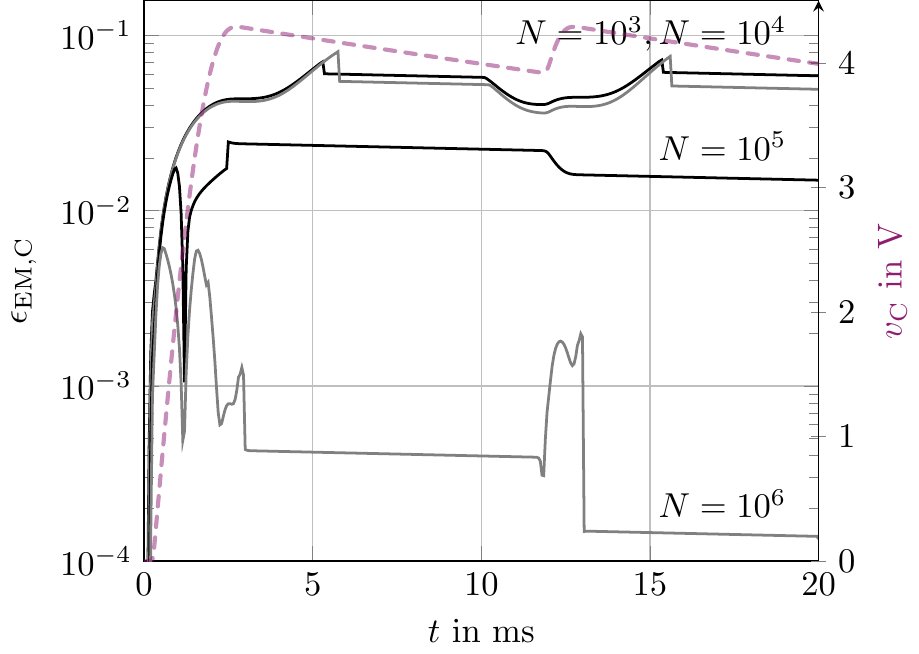}
		\caption{Half-wave rectifier: Energy mismatch at the capacitor C over time. The error is shown for an increasing number of measurement data $N \in \left\{10^3, 10^4, 10^5, 10^6\right\}$. The dashed line shows the output voltage $v_\mathrm{C}$.}
		\label{fig:half-wave-rectifier_uC_EM_over_time}
\end{center}
\end{figure}

\begin{figure}[t!]
	\begin{center}
		\includegraphics[scale=0.80]{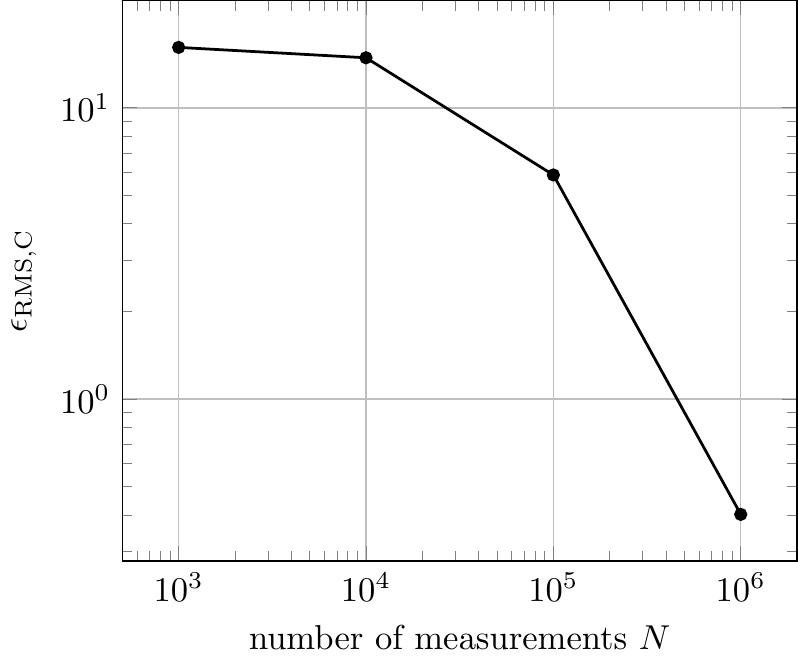}
		\caption{Half-wave rectifier: \gls{rms} error at the capacitor C over number of measurements.}
		\label{fig:half-wave-rectifier_rmsC}
	\end{center}
\end{figure}

We assume that the resistors and the capacitor are known elements, whereas the diode is only known by measurement data. The parasitic resistor $R_\mathrm{D}$ is part of the circuit model in the \textsc{ngspice} simulation \cite{nenzi2011ngspice}. Hence, the measurement data is generated from the diode model including the parasitic resistor $R_\mathrm{D}$. The data-driven solutions are compared to a reference solution obtained with \textsc{ngspice}. 
For both the reference and the data-driven solutions, $K=400$ time steps are used. The data-driven solver is based on the trapezoidal rule. Figure~\ref{fig:half-wave-rectifier_uC} shows the output voltage $v_\mathrm{C}$ of the half-wave rectifier. The figure clearly shows that the data-driven solution becomes more accurate if more measurement data is employed. However, we also observe that a rather large number of measurement data is needed to achieve an adequately accurate solution. This behavior can also be seen in Figure~\ref{fig:half-wave-rectifier_uC_EM_over_time}, which shows the energy mismatch over time, as well as in Figure~\ref{fig:half-wave-rectifier_rmsC}, which shows the \gls{rms} error at the capacitor. The reduced convergence rate can be explained by the switching behavior of the diode. That is, the distance-minimizing data-driven solver faces problems in the regime where the current of the diode remains almost constant, i.e., for $v_\mathrm{D} \ll v_\mathrm{fv}$, where $v_\mathrm{fv}$ is the forward threshold voltage.

%% file: computational_costs.tex
In the following, we briefly discuss the computational cost of the data-driven \gls{mna} solver. For a more detailed discussion on the computational complexity of data-driven solvers, the reader is referred to \cite{galetzka2020datadriven, galetzka2021datadriven}. 

In each data-driven iteration, we have to solve a linear system of size $2n-2 + n_\mathrm{G}+n_\mathrm{C}+3n_\mathrm{L}+2n_\mathrm{V}$. The number of data-driven iterations is strongly connected to the number of employed measurement data. Furthermore, strongly nonlinear element responses also demand for more data-driven iterations. In the case of the linear RC-circuit, the number of data-driven iterations per time step vary from $5$ (for $N=10^2$) to $45$ (for $N=10^6$). For the nonlinear RC-circuit, the number of iterations depends on the current operation point in the $(u_\mathrm{C},q_\mathrm{C})$ phase-space. In the region where the relation is almost linear, the number of data-driven iterations is similar to the ones of the linear RC-circuit. If the operation point is in the strongly nonlinear regions, the iterations range from $20$ (for $N=10^2$) to $140$ (for $N=10^6$). In comparison, the traditional solver must solve the nonlinear problem at each time step, which 
resulted in $12$ iterations on average in the case of the nonlinear RC circuit. 

Hence, the data-driven \gls{mna} solver in its current implementation still needs up to ten times more simulation time. This is the price to pay for the benefit of not requiring models for the circuit elements. It is expected that progress in data-processing will further decrease the gap between traditional and data-driven \gls{mna} solvers.

Last, we note that the aforementioned remarks on the  computational cost of the traditional and the data-driven \gls{mna} solvers only take into account the aspect of computational complexity in terms of linear system size, system solutions, and solver iterations.
That is, this discussion does not include the effort that is usually needed to identify and construct model representations to be used within traditional \gls{mna} solvers.  While this element modeling effort is difficult to assess, it must be underlined that the modeling step is bypassed altogether when the data-driven solver is employed. 

%% file: conclusion.tex
This paper proposes a data-driven \gls{mna} circuit solver, which relies on Kirchhoff's laws and on measurement data for the elements instead of traditional element models.
The solver was first applied to a linear RC-circuit, where an investigation on the different approximation errors concluded that the available measurement data and the time step employed in the time integration method should match one another in order to avoid unnecessary computational costs. The solver was additionally applied to a nonlinear RC-circuit with a voltage-dependent capacitor and to a half-wave rectifier with a data-driven diode. In the former case, the solver achieves a linear convergence rate with respect to the number of measurement data points. This convergence rate is not observed in the latter case, where the switching behavior of the diode model, in particular the regime before the threshold voltage, hinders solver convergence and demands for comparatively large data sets.

The numerical results have shown that the data-driven \gls{mna} solver developed in this work constitutes a viable alternative for standard \gls{mna} circuit simulation, in particular for cases featuring elements with an unknown model representation, but for which measurement data are available instead.
Due to the identified limitations of the data-drive \gls{mna} solver in handling circuits with a switching behavior, such as the half-wave rectifier unit shown in Section~\ref{subsec:rectifier}, future research efforts shall concentrate on a more in-depth investigation of these specific cases. 
Additional topics for future research include data-driven circuit simulation incorporating noisy measurement data, as well as the handling of elements that cannot be described through the classical two-port network model.